\documentclass{JHEP3}

\usepackage{amssymb}
\usepackage{amsmath}

\usepackage{graphicx}
\usepackage{graphics}
\usepackage{epsfig}

\usepackage{multirow}

\newcommand{\APJ}{Astrophys.\ J.\/}

\newcommand{\MNRAS}{Mon.\ Not.\ R.\ Aston.\ Soc.\/} 
\newcommand{\NP}{Nucl.\ Phys.\/} 

\newcommand{\PL}{Phys.\ Lett.\/} 
\newcommand{\PR}{Phys.\ Rev.\/} 
\newcommand{\PRL}{Phys.\ Rev.\ Lett.\/} 
\newcommand{\lcdm}{$\Lambda$CDM~}

\title{Dynamical Dark Energy model parameters with or without massive
neutrinos.}

\author{G. La Vacca$^{1,2}$, J.R. Kristiansen$^3$\\
$^1$Physics Department G.~Occhialini, Milano--Bicocca
University, Piazza della Scienza 3, 20126 Milano, Italy\\
$^2$I.N.F.N., Sezione di Milano--Bicocca, Piazza della
Scienza 3, 20126 Milano, Italy\\
$^3$Institute of Theoretical Astrophysics, University of Oslo,
Box 1029, 0315 Oslo, Norway\\
E-mail: \email{lavacca@mib.infn.it,j.r.kristiansen@astro.uio.no}}

\abstract{We use WMAP5 and other cosmological data to constrain model
parameters in quintessence cosmologies, focusing also on their shift
when we allow for non--vanishing neutrino masses. The Ratra--Peebles
(RP) and SUGRA potentials are used here, as examples of slowly or
fastly varying state parameter $w(a)$. Both potentials depend on an
energy scale $\Lambda$. Here we confirm the results of previous
analysis with WMAP3 data on the upper limits on $\Lambda$, which turn
out to be rather small (down to $\sim 10^{-9}$ in RP cosmologies and
$\sim 10^{-5}$ for SUGRA). Our constraints on $\Lambda$ are not
heavily affected by the inclusion of neutrino mass as a free
parameter. On the contrary, when the neutrino mass degree of freedom
is opened, significant shifts in the best--fit values of other
parameters occur.}

\preprint{...}

\begin{document}

\section{Introduction}
\label{sec:intro}

Recent data require an accelerated expansion of the Universe.
High-redshift supernovae~\cite{sn}, CMB~\cite{cmb} and deep
samples~\cite{lss} essentially agree with requiring a density
parameter for non--relativistic matter $\Omega_m \sim 0.27$, while the
total density parameter $\Omega_0 \sim 1$. A fluid, dubbed Dark Energy
(DE), filling the gap in the cosmological energy balance, can account
for cosmic acceleration if its state parameter $w \equiv
\rho_{de}/P_{de}$ is close to -1.

In the minimal model, dubbed \lcdm (or {\it cosmic concordance})
cosmology, DE has a state parameter $w \equiv -1$ as though being a
false vacuum, and this is equivalent to introducing Einstein's
cosmological constant. All data available up to now can be accomodated
in this cosmology. Still there are several problems related to the
nature of the DE. Infact, two critical issues plague such model: the
{\it fine tuning} of the present DE energy density to the vacuum
expectation value of quantum field theories, and the {\it coincidence}
between DE domination and the advent of the non--linear growth of
matter fluctuations.

In the attempt to approach a better understanding of DE nature, a
varying state parameter $w(a)$ has been tentatively allowed. Here two
patterns can be followed: either phaenomenological fitting expressions
for $w(a)$ can be introduced or, if we assume that DE is a
self--interacting scalar field $\phi$ \cite{quintessence}, {\it
tracking} potential expressions $V(\phi)$~\cite{tracking}, depending
on suitable parameters, can be tested. This latter approach has the
advantage of allowing for larger energy scales of DE, thus easing the
fine tuning problem.

Analysis of such dynamical Dark Energy (dDE) scenario were previously
performed when WMAP3 data became available \cite{bib6}. The first aim
of this work is to inspect how far their output are modified when
WMAP5 data are considered, in analogy to the modifications occurring
for \lcdm cosmologies. As in \cite{bib6}, we shall do so by using RP
and SUGRA potentials (see below).

Here we shall however further focus on the impact on parameters
arising from allowing for a neutrino mass. As a matter of fact,
cosmology is sensitive to $\nu$ masses (see \cite{bonovalda},
\cite{lesgourgues:2006}) even at a level below current limits set by
$\beta$--decay experiments \cite{bdecay}, so much that the very
opening of the degree of freedom of neutrino mass, causes appreciable
shifts in the best--fit values of some cosmological parameters. It
should be outlined that non--vanishing neutrino masses are required by
flavor mixing results \cite{deltam}. Even though such results set
quite a low limit to the sum $M_\nu$ of neutrino masses, it would be
therefore unappropriate to neglect the neutrino mass option.

This fact was recently exploited by several researchers to show that a
larger coupling between CDM and DE is compatible with data
\cite{lavacca,kristiansen,catalans} when allowing for a non-zero
neutrino mass. Here, however, no such coupling shall be considered.

Let us finally outline that further alternatives to a cosmological
constant or a scalar field have been considered. Among them,
modifications to Hilbert--Einstein Lagrangian density, adding $f(R)$
terms ($R$: Ricci scalar), or the hypothesis that DE arises as a back
reaction to the development of inhomogeneities \cite{backreaction}.
Such options make the need of an extra {\it substance} superfluous,
but give rise to other theoretical and observational problems. We
shall not deal with them in this work.

The plane of the paper is as follow. In Sec. \ref{sec:models} we
describe the models. Data and methods used are reviewed in
Sec. \ref{sec:methods}, while in Sec. \ref{sec:results} we present
results and discuss our findings.

\section{Models}
\label{sec:models}

In a dDE scenario and in the reference frame where the metric is $ds^2
= a^2(\tau)[d\tau^2 - d\ell^2]$ ($\tau:$ conformal time; $d\ell:$
spatial line element), the energy density and pressure of the field
$\phi$ read
\begin{equation}
\rho_{de} = \rho_k + V(\phi)~,~~~ P_{de} = \rho_k -
V(\phi)~,~~~~~~~~~~~ {\rm with}~~\rho_k = (\phi')^2/2a^2~,
\label{rhop}
\end{equation}
prime indicating differentiation in respect to $\tau$. Accordingly, if
$\rho_k \gg V$, the DE state parameter approaches +1 ({\it stiff
matter}) so that DE energy density rapidy dilutes during expansion
($\rho \propto a^{-6}$). In the opposite case $V \gg \rho_k$, the
state parameter approaches $-1$ and DE is suitable to explain the
observed cosmic acceleration.

The equations
\begin{equation}
\rho'_{de} + 3\frac{a'}{a} (P_{de} + \rho_{de}) = 0,
\end{equation}
and
\begin{equation}
\label{eom}
\phi'' + 2\frac{a'}{a} \phi' + a^2 V,_\phi = 0
\end{equation}
are then clearly equivalent. By integrating (\ref{eom}) together with
the Friedmann equations, we obtain the time evolution of $\phi$ and,
thence, of $\rho_k$, $V(\phi)$, $\rho_{de}$, $P_{de}$ and, finally,
$w(a)$.

Quite in general, the solution of a differential equation depends on
initial conditions. {\it Tracker potentials}, however, have attractor
solutions ({\it tracking solutions}) on which they converge (almost)
independently of the initial conditions.

\begin{figure}[!t]
\begin{center}
\includegraphics[height=7.cm,angle=0]{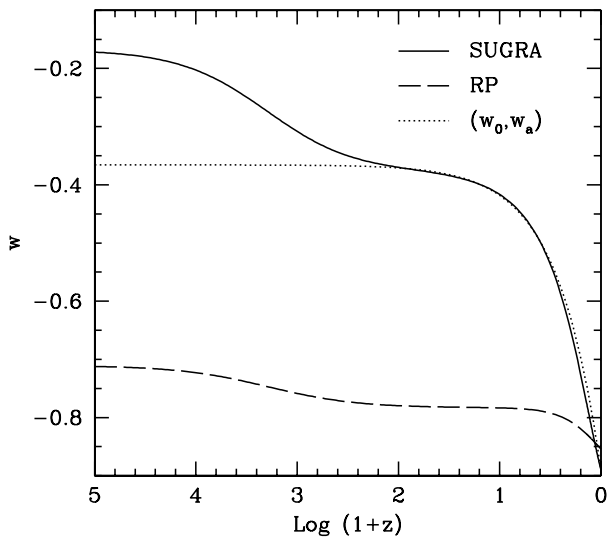}
\end{center}
\caption{Comparison between the variation of the state parameter for
RP and SUGRA potentials and the usual parametrization $w(a) = w_o +
(1+a)w_a$.  Here, $\Lambda$ is 100~GeV for SUGRA and $10^{-8}$~GeV for
RP, while $(w_o,~w_a)$ = (-0.8908, 0.525). These values are selected
to yield a behavior close to SUGRA, at low z. Even renouncing to a
full coincidence at $z=0$, the fast variablility of $w(a)$ in SUGRA
cannot be met by any polinomial $w(a)$. Notice also that, for a RP
potential, $w(a)$ varies more slowly. The plot is for $h=0.7$,
$\Omega_b=0.046$, $\Omega_c= 0.209$.}
\label{fig:SU}
\end{figure}

Among tracker potentials, we consider here:
\begin{align}
V(\phi)& = \Lambda^{\alpha+4}/\phi^\alpha && {\rm RP}~\cite{RP} \\
V(\phi)& = (\Lambda^{\alpha+4}/\phi^\alpha) \exp(4\pi\, \phi^2/m_p^2)
&&{\rm SUGRA}~\cite{sugra}.
\end{align}
The RP potential yields a slowly varying $w(a)$ state parameter, while
the variation of $w(a)$ for SUGRA potential is faster, as shown in
Figure \ref{fig:SU}.  
In both cases, for any choice of $\Lambda$ and $\alpha$ these
potentials yield a precise DE density parameter $\Omega_{de}$. Here we
use $\Lambda$ and $\Omega_{de}$ as free parameters in flat
cosmologies; the related $\alpha$ value is then suitably fixed.

Let us then remind the constraints on RP and SUGRA potentials obtained
when using WMAP3 data \cite{bib6}. They confirmed that a RP cosmology
requires an energy scale $\Lambda$ below $\sim 10^{-7}{\rm GeV}$ (and
then $\alpha \sim 0.5$); this is coherent with the fact that $w(a)$ is
then steadily below -0.85 ($ \sim 95\, \% $~C.L.)  as required from
observations for a constant $w$ \cite{komatsu}. On the contrary, in
the SUGRA case, while their best fit still leads to low $\Lambda$'s,
they found that a value $\Lambda \sim$~100~GeV is within $\sim 2\,
\sigma$'s from the best--fitting model; such scale is then compatible
with the ones of SUSY or EW transition.

One significant result of our analysis is that these upper limits are
badly lowered when using WMAP5 data.

\section{Methods}
\label{sec:methods}

For this work we use a modified version of the cosmological Boltzmann
code CAMB \cite{Lewis:1999bs, camb:2008} to calculate theoretical CMB
and matter power spectra. 
%
In our modified code, a precursor program is called before running
CAMB, evolving the background solutions in order to determine
iteratively the value of $\alpha$ consistent with the assigned
$\Omega_{de}$ and $\Lambda$. The scalar field $\phi$ evolves according
to the Klein--Gordon equation (\ref{eom}). Its initial value $\phi_0$
is obtained considering that $\rho_{de,0}\sim V(\phi_0)$. In the code
we also took into account spatial fluctuations of the DE field.

Our code is embedded into the publicly available Monte Carlo Markov
Chain (MCMC) engine CosmoMC \cite{lewis:2002}. The following set of
free parameters are used: $\{ \omega_b,\, \omega_{c},\, \theta,\,
\tau_o,\, n_s,$ $\, \ln (10^{10} A_s), \, \lambda,\, M_\nu\}$. Here
$\omega_{b,c}= \Omega_{b,c} h^2$ are the reduced density parameters
for baryonic and cold dark matter, respectively, $h$ being the present
value of the Hubble constant in units of 100 km/s/Mpc; $\theta$ is
then the ratio of the sound horizon to the angular diameter distance
at recombination, $\tau_o$ is the optical depth, $n_s$ is the scalar
spectral index, $A_s$ is the primordial amplitude of scalar
fluctuations at a scale $k=0.05 \textrm{Mpc}^{-1}$ and
$\lambda\equiv\log_{10}(\Lambda/{\rm GeV})$. In addition we
marginalize over the SZ amplitude.  We assume three massive neutrino
species with degenerate masses.

In the MCMC runs we combined the five year data from the WMAP
measurements of the CMB radiation (WMAP5) with the galaxy power
spectrum from the 2dF survey \cite{cole:2005}, supernovae type 1a data
from the SNLS survey \cite{astier:2006}, and added the priors: $h=0.72
\pm 0.08$ (from the HST key project \cite{freedman:2001}), $\omega_b =
0.022 \pm 0.002$ (from BBN results \cite{burles:2001, cyburt:2004,
serpico:2004}).

\section{Results and discussion}
\label{sec:results}

In Table \ref{tab:params} we report best fit parameter values and
1--$\sigma$ errors for different models. In the case of neutrino
masses, whose sum $M_\nu$ is not determined by data, we give its
(95$\, \%$ C.L.)  upper limit. Also for the scales $\lambda =
\log(\Lambda/ {\rm GeV})$, in the RP and SUGRA potentials, an upper
limit is given.
\begin{table}[!t]
\begin{center}
\begin{tabular}{cllllll}
\hline
\multirow{2}{*}{Parameter}& $\Lambda$CDM & $w=$ const. & \multicolumn{2}{c}{RP} & \multicolumn{2}{c}{SUGRA} \\
 & $M_\nu \neq 0$ & $M_\nu \neq 0$ & $M_\nu \equiv 0$ & $M_\nu \neq 0$ & $M_\nu \equiv 0$ & $M_\nu \neq 0$\\
\hline
\\
\multirow{2}{*}{$10^2\, \omega_{b}$} 
& 2.258       & 2.247       & 2.278        & 2.271        & 2.278        & 2.274\\
& $\pm$ 0.061 & $\pm$ 0.062 & $\pm$ 0.060  & $\pm$ 0.060  & $\pm$ 0.060  & $\pm$ 0.060 \\
\\
\multirow{2}{*}{$\omega_{c}$}        
& 0.1098       & 0.1132       & 0.1051       & 0.1062       & 0.1043       & 0.1055\\
& $\pm$ 0.0040 & $\pm$ 0.0069 & $\pm$ 0.0050 & $\pm$ 0.0047 & $\pm$ 0.0051 & $\pm$ 0.0050 \\
\\
\multirow{2}{*}{$H_o$ {\rm (km/s/Mpc)}} 
& 70.1      & 69.7      & 70.9        & 69.8        & 70.6        & 69.6 \\
& $\pm$ 2.1 & $\pm$ 2.2 & $\pm$ 1.7   & $\pm$ 2.0   & $\pm$ 1.9   & $\pm$ 2.0 \\
\\
\multirow{2}{*}{$\tau_o$}              
& 0.087       & 0.085       & 0.089        & 0.090        & 0.089        & 0.091\\
& $\pm$ 0.017 & $\pm$ 0.017 & $\pm$ 0.017  & $\pm$ 0.018  & $\pm$ 0.017  & $\pm$ 0.018 \\
\\
\multirow{2}{*}{$w$}              
& \multirow{2}{*}{-1} & -1.06       & \multirow{2}{*}{--}  & \multirow{2}{*}{--} & \multirow{2}{*}{--} & \multirow{2}{*}{--}\\
&                     & $\pm$ 0.10 \\
\\
$M_\nu$ (eV) 
& \multirow{2}{*}{$<$ 0.66} & \multirow{2}{*}{$<$ 0.94}& \multirow{2}{*}{--} 
& \multirow{2}{*}{$<$ 0.58} & \multirow{2}{*}{--} & \multirow{2}{*}{$<$ 0.59}   \\
(95\% C.L.)
\\
$\lambda$ 
& \multirow{2}{*}{--} & \multirow{2}{*}{--} & \multirow{2}{*}{$<$ -8.4} 
& \multirow{2}{*}{$<$ -8.7} & \multirow{2}{*}{$<$ -3.4} & \multirow{2}{*}{$<$ -4.7} \\   
(95\% C.L.)
\\
\multirow{2}{*}{$n_s$}                  
& 0.962       & 0.958       & 0.968       & 0.967       & 0.970       & 0.967 \\
& $\pm$ 0.014 & $\pm$ 0.015 & $\pm$ 0.014 & $\pm$ 0.014 & $\pm$ 0.014 & $\pm$ 0.014 \\
\\
\multirow{2}{*}{{\rm ln}$(10^{10}A_s)$} 
& 3.045       & 3.049       & 3.047       & 3.041       & 3.045       & 3.041\\
& $\pm$ 0.040 & $\pm$ 0.040 & $\pm$ 0.041 & $\pm$ 0.042 & $\pm$ 0.041 & $\pm$ 0.041 \\
\\
\multirow{2}{*}{$\Omega_m$}             
& 0.270       & 0.280       & 0.254       & 0.266       & 0.255       & 0.266       \\
& $\pm$ 0.022 & $\pm$ 0.027 & $\pm$ 0.017 & $\pm$ 0.021 & $\pm$ 0.018 & $\pm$ 0.020 \\
\\
\multirow{2}{*}{$\sigma_8$}             
& 0.713       & 0.711       & 0.749       & 0.697       & 0.737       & 0.689\\
& $\pm$ 0.056 & $\pm$ 0.059 & $\pm$ 0.039 & $\pm$ 0.052 & $\pm$ 0.046 & $\pm$ 0.054 \\
\\
\hline
-2 ln($\cal L$) & 1407.25 & 1407.38 &  1407.35  &  1407.38   & 1407.44  &  1407.31 \\

\end{tabular}
\caption{Constraints on \lcdm and $w=$const models, with massive
$\nu$'s, and RP and SUGRA models, with and without massive
$\nu$'s. Limits obtained using the combination of data sets described
in the text. The last row gives the likelihood of the best-fit sample
for each model.}
\label{tab:params}
\end{center}
\end{table}

Let us first comment on the \lcdm and $w=$ const. column, which was
added here for the sake of comparison. In both of them neutrino masses
are allowed. As expected, the values we find for all parameters are
strictly coeherent with those found by the WMAP team when combining
the WMAP5 data with measurements of the position of the baryonic
acoustic oscillation (BAO) peak in the galaxy power spectrum and SNLS
data. The main discrepancy concerns error bars, being systematically
slightly greater here than in the outputs from the WMAP team.

The reason for this can be easily understood. Deep sample data as 2dF
and SDSS are substantially coherent, as far as the BAO location is
concerned. Some discrepancy between them is however present in the
spectral details.  Also because of that, the WMAP5 fit takes into
account the BAO position only. On the contrary, here we took into
account the whole information contained in the 2dF spectral data.  In
addition we have used slightly different priors than the WMAP team
did, so that the (small) discrepancies found are fully justified. In
particular, our results for a constant $w=-1$ are largely consistent
with the analysis of the WMAP team.

Infact, for $w=-1$ we find an upper limit of $M_\nu < 0.66$eV, while
WMAP5+SNLS+BAO yields $M_\nu < 0.61$eV.  However, when $w \neq -1$
models are considered, the upper limit exhibits a sharp increase, up
to 0.94~eV. WMAP5+SNLS+BAO outputs show only a $\sim 10\, \%$ shift in
the upper limit when passing from \lcdm to $w \neq -1$.  Here the
shift is $\sim 4$ times greater. This is related to the widening of
the $w$ interval, as greater $M_\nu$ values become compatible with
data if we delve into the {\it phantom} regime
\cite{Hannestad:2005gj}. Accordingly, as RP and SUGRA models yield an
effective equation of state $w(a) > -1$, this leads to more stringent
limits on $M_\nu$.

As far as RP and SUGRA models are concerned, the decrease of errors in
respect to WMAP3 outputs in \cite{bib6} has direct consequences on the
limits on the energy scale $\Lambda$. In the case $M_\nu \equiv 0$,
treated in \cite{bib6}, an upper limit $\lambda < -7.7$ ($2.1$) was
found for RP (SUGRA) at 95$\, \%$ C.L.. Such limits are now lowered to
-8.4 for RP and to a severe -3.4 in the case of SUGRA. Limits are even
more restrictive if $M_\nu \neq 0$ is allowed.

\begin{figure}[htb]
\begin{center}
\includegraphics[height=8.cm,angle=0]{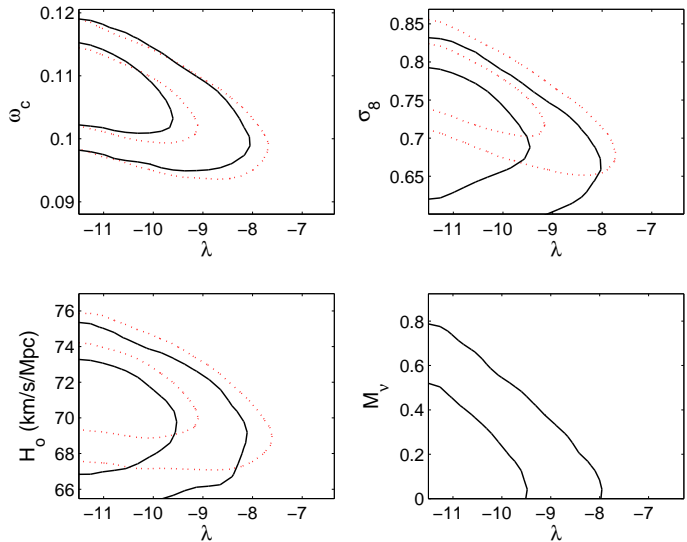}
\end{center}
\caption{2D contour plots with $\lambda$ vs $\omega_c$, $\sigma_8$,
$H_0$ and $M_\nu$ for models with a RP potential. Solid black lines
are for models free $M_\nu$, and dotted red lines show the contours
for models with $M_\nu = 0$}
\label{fig:RP_2D}
\end{figure}

\begin{figure}[htb]
\begin{center}
\includegraphics[height=8.cm,angle=0]{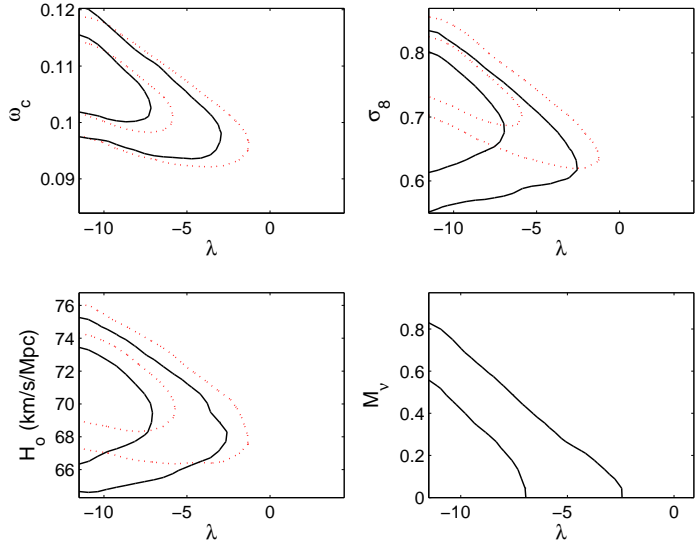}
\end{center}
\caption[]{Same as Figure \ref{fig:RP_2D}, but with a SUGRA potential.}
\label{fig:SUGRA_2D}
\end{figure}

The reason for this strong improvement in the limits on $\lambda$ can
partly be understood by inspecting Figures \ref{fig:RP_2D} and
\ref{fig:SUGRA_2D}. Here we show 2D contours for $\lambda$ versus
$\omega_c$, $\sigma_8$, $H_0$ and $M_\nu$, which are parameters that
are strongly correlated with $\lambda$. In the plots we compare the
results from the models with $M_\nu = 0$, to those with a free
neutrino mass. There is, for example, strong anticorrelations between
$\sigma_8$ and $\lambda$ and $\omega_c$ and $\lambda$. As our analysis
gives a preference of significantly larger $\omega_c$ and $\sigma_8$
than the WMAP3 analysis in \cite{bib6}, we also find much tighter
limits on $\lambda$. Parts of the improved limits can also be
attributed to our general decrease in error bars on the parameters.


As far as the other parameters are concerned, all shifts in respect to
\lcdm are within 1--$\sigma$. It may be however worth considering the
2--$\sigma$ upper limit for $n_s$. When using WMAP3 data there was a
discrepancy between the \lcdm and the RP or SUGRA cases. In the former
case $n_s = 1$ was excluded at the 95$\, \%$ C.L.; on the contrary,
$n_s = 1$ kept consistent with data, at the same C.L., for RP and
SUGRA. Using WMAP5 data the discrepancy disappears and $n_s = 1$ is
above the 95$\, \%$ C.L. for all models considered. This means that
the related constraint on inflationary models is apparently much more
independent from DE nature.

Another significant shift worth outlining concerns $\sigma_8$ values.
Let us remind that
\begin{equation}
\sigma_8^2 = {A_s \over 2 \pi^2} \int dk~k^{2 + n_s} {\cal T}^2 (k)
W^2(kR_8)~.
\label{s8}
\end{equation}
Here: $\cal T$$(k)$ is the transfer function, $W(kR)$ is the Fourier
transform of a (top--hat) filter, $R_8 = 8\, h^{-1}$Mpc~.

Although in the absence of major shifts on $A_s$ and $n_s$, it is
sufficient to open the degree of freedom of neutrino mass, to allow
for significant shifts in $\cal T$ which, in turn, reflect onto
a smaller best--fit $\sigma_8$ value.

Also likelihood values for the best--fitting models do not exhibit
major changes and cannot be used to discriminate among the cosmologies
considered here. This is coherent with the general conclusion we draw,
that current data are still unable favor any kind of DE nature.

To summarize, we have studied dDE models with RP and SUGRA potentials,
and used a compilation of cosmological data sets to constrain
parameters of these models. We have also studied the effect of opening
an extra degree of freedom, by allowing for non-zero neutrino masses.

Our basic result is that the energy scale of the DE model can be
constrained to $\lambda < -8.4$ (-3.4) if $M_\nu \equiv 0$ and to
$\lambda < -8.7$ (-4.7) in presence of massive $\nu$, when using a RP
(SUGRA) potential. These limits are significantly stronger than what
was found in \cite{bib6}, where they used WMAP3 data and vanishing
neutrino masses only. Some of the improvement can be attributed to the
inclusion of neutrino masses, as non-zero neutrino masses tend to
shrink the allowed parameter space for $\lambda$ (see Figures
\ref{fig:RP_2D} and \ref{fig:SUGRA_2D}). However, the most of the
improvement is caused by our use of more precise data, and that these
data have a preference of slightly different values of parameters like
$\sigma_8$ and $\omega_c$, which in turn drives $\lambda$ to smaller
values.

\acknowledgments 

It is a pleasure to thank Silvio Bonometto for helpful
discussions. Thanks are also due to Loris Colombo and Roberto Mainini
for the use of the modified version of CAMB and CosmoMC. JRK
acknowledges support from the Research Council of Norway. This work
was partially supported by ASI (the Italian Space Agency) thanks to
the contract I/016/07/0 "COFIS". The computations presented in this
paper were carried out on Titan, a cluster owned and maintained by the
University of Oslo and NOTUR.


\begin{thebibliography}{50}

\bibitem{sn} 
  Riess A G, Kirshner R P, Schmidt B P, Jha S, et al. 1998 \APJ {\bf 116} 1009;
  Perlmutter S, Aldering G, Goldhaber G, et al. 1999 \APJ {\bf 517} 565; 
  Riess A G, Strolger L-G, Tonry J, et al. 2004 \APJ {\bf 607} 665
\bibitem{cmb} 
  de~Bernardis P, Ade P A R, Bock J J, et al. 2000 {\it Nature} {\bf 404} 955; 
  Padin S, Cartwright J K, Mason B S, et al. 2001 \APJ {\bf 549} L1; 
  Kovac J, Leitch E M, Pryke C, et al. 2002 {\it Nature} {\bf 420} 772; 
  Scott P F, Carreira P, Cleary K, et al. 2003 \MNRAS {\bf 341} 1076; 
  Spergel D N, Bean R, Dor{\`e} et al. 2007 \APJ {\bf 170} 377 {\it Preprint} 
   astro--ph/0603449
\bibitem{lss} 
  Colless M M, Dalton G B, Maddox S J, et al. 2001 \MNRAS {\bf 329} 1039; 
  Colless M M, Peterson B A, Jackson C, et al. 2003 {\it Preprint} astro--ph/0306581; 
  Loveday J (the SDSS collaboration) 2002 {\it Contemporary Phys.} {\bf 43} 437; 
  Tegmark M, Blanton M, Strauss M et al. 2004 \APJ {\bf 606} 702--740; 
  Adelman--McCarthy J K, Agueros M A, Allam S S, et al. 2006 {\it Astrophys. J. Suppl.} {\bf 162} 38--48
\bibitem{quintessence} 
  Caldwell R R, Dave R and Steinhardt P J 1998 \PRL {\bf 80} 1582; 
  Wetterich C 1988 \NP B{\bf 302} 668
\bibitem{tracking} 
  Zlatev I, Wang L and Steinhardt P J 1999 \PRL {\bf 82} 896--899; 
  Steinhardt P J, Wang L and Zlatev I 1999 \PR D{\bf 59} 123504
\bibitem{bib6} 
  L. Colombo \& M. Gervasi, JCAP {\bf 10}, 001 (2006).
\bibitem{bonovalda}
  S.A.~Bonometto \& R.~Valdarnini, \PL A {\bf 104}, 369 (1984);
  R. Valdarnini \& S. Bonometto, A\&A {\bf 146}, 235 (1985);
  R. Valdarnini \& S. Bonometto, \APJ {\bf 299}, L71 (1985);
  Achilli S., Occhionero F. \& Scaramella R., \APJ {\bf 299}, 577 (1985)
\bibitem{lesgourgues:2006}
  J. Lesgourgues \& S. Pastor, Phys. Rept. 429 (2006) 307-379
\bibitem{bdecay}
  E W Otten and C Weinheimer 2008 Rep. Prog. Phys. {\bf 71} 086201;
  H.~V.~Klapdor-Kleingrothaus, I.~V.~Krivosheina, A.~Dietz and O.~Chkvorets, \PL B {\bf 586} (2004) 198;
  H.~V.~Klapdor-Kleingrothaus, arXiv:hep-ph/0512263;
  C.~Arnaboldi {\it et al.}  [CUORICINO Collaboration], Phys.\ Rev.\  C {\bf 78}, 035502 (2008);
  Bakalyarov {\it et~al.}, [GERDA collaboration], 2008, http://www.mpi-hd.mpg.de/gerda/reportsLNGS/gerda-lngs-sc-oct08-shwup.pdf;
  KATRIN Collaboration, 2005, http://www-ik.fzk.de/~katrin/publications/documents/DesignReport2004-12Jan2005.pdf
\bibitem{deltam}
  Q. R.Ahmad {\it et al.}, \PRL {\bf 89}, 011301 (2002);
  S. N.Ahmed {\it et al.}, \PRL {\bf 92}, 181301 (2004);
  K. Eguchi {\it et al.}, \PRL {\bf 90}, 021802h (2003); 
  T. Araki {\it et al.}, \PRL {\bf 94}, 081801 (2005);
  W. W. Allison {\it et al.}, \PL B{\bf 449}, 137 (1999); 
  M. Ambrosio {\it et al.}, \PL B{\bf 517}, 59 (2001);
  M. H. Ahn {\it et al.}, \PRL {\bf 90}, 041801h (2003);
  D. G. Michael {\it et al.}, \PRL {\bf 97}, 191801 (2006).
\bibitem{lavacca}
  G.~La Vacca, S.~A.~Bonometto and L.~P.~L.~Colombo, 2009, New Astron. 14, 435, arXiv:0810.0127 [astro-ph];
  G.~La Vacca {\it et al.}, arXiv:0902.2711 [astro-ph.CO].
\bibitem{kristiansen}
  J.~R.~Kristiansen {\it et al.}, arXiv:0902.2737 [astro-ph.CO].
\bibitem{catalans}
  M.~B.~Gavela {\it et al.}, arXiv:0901.1611 [astro-ph];
\bibitem{backreaction}
  Buchert T., Kerscher M. \& Sicka C., 2000, Phys.Rev., D62, 043525;
  Buchert T., 2000, Gen.Rel.Grav., 32, 105;
  Kolb E.W., Matarrese S. \& Riotto A., 2005, astro-ph/0506534;
  Kolb E.W., Matarrese S., Notari A. \& Riotto A., 2005,  Phys.Rev., D71, 023524
\bibitem{RP} 
  Ratra B. \& Peebles P.J.E. (1988) PR D{\bf 37}, 3406.
\bibitem{sugra} 
  Brax P and Martin J 1999 \PL B{\bf 468} 40; 
  Brax P and Martin J 2000 \PR D{\bf 61} 103502
\bibitem{komatsu} 
  E. Komatsu {\it et al.} [WMAP Collaboration], arXiv:0803.0547 [astro-ph].
\bibitem{Lewis:1999bs}
  A.~Lewis, A.~Challinor and A.~Lasenby 2000 \APJ {\bf 538} 473
\bibitem{camb:2008}
  http://camb.info
\bibitem{lewis:2002}
  A.~Lewis and S.~Bridle 2002 \PR D {\bf 66}, 103511
\bibitem{cole:2005} 
  S.~Cole {\it et al.}  [The 2dFGRS Collaboration] 2005 \MNRAS {\bf 362} 505
\bibitem{astier:2006}
  P.~Astier {\it et al.}  [The SNLS Collaboration] 2006  Astron.\ Astrophys.\  {\bf 447}, 31
\bibitem{freedman:2001} 
  W.~L.~Freedman {\it et al.}  [HST Collaboration] 2001 Astrophys.\ J.\  {\bf 553} 47
\bibitem{burles:2001}
  S.~Burles, K.~M.~Nollett and M.~S.~Turner 2001 \PR D {\bf 63} 063512
\bibitem{cyburt:2004}
  R.~H.~Cyburt 2004 Phys.\ Rev.\  D {\bf 70}, 023505
\bibitem{serpico:2004}
  P.~D.~Serpico, S.~Esposito, F.~Iocco, G.~Mangano, G.~Miele and O.~Pisanti 2004 JCAP {\bf 0412} 010
\bibitem{Hannestad:2005gj}
  S.~Hannestad, \PRL {\bf 95}, 221301 (2005).
\bibitem{DeLaMacorra:2006tu}
  A.~De La Macorra, A.~Melchiorri, P.~Serra \& R.~Bean, Astropart.\ Phys.\  {\bf 27}, 406 (2007).

\end{thebibliography}
\end{document}